# Comparative Study of Planetary Atmospheres and Implications for Atmospheric Entry Missions

Athul Pradeepkumar Girija [1,****]

[1]*School of Aeronautics and Astronautics, Purdue University, West Lafayette, IN 47907, USA*

## ABSTRACT

The study of planetary atmospheres is critical to our understanding of the origin and evolution of the Solar System. The combined effect of various physical and chemical processes over billions of years have resulted in a variety of planetary atmospheres across the Solar System. This paper performs a comparative study of planetary atmospheres and their engineering implications for future entry and aerocapture missions. The thick Venusian atmosphere results in high deceleration and heating rates and presents a demanding environment for both atmospheric entry and aerocapture. The thin Martian atmosphere allows low aerodynamic heating, but itself is not enough to decelerate a lander to sufficiently low speeds for a soft landing. With their enormous gravity wells, Jupiter and Saturn entry result in very high entry speeds, deceleration, and heating making them the most demanding destinations for atmospheric entry and impractical for aerocapture. Titan is a unique destination, with its low gravity and greatly extended thick atmosphere enabling low deceleration and heating loads for entry and aerocapture. Uranus and Neptune also have large gravity wells, resulting in high entry speeds, high deceleration and heating compared to the inner planets, but are still less demanding than Jupiter or Saturn.

***Keywords:*** Planetary Atmosphere, Comparative Study, Atmospheric Entry

---

[****] To whom correspondence should be addressed, E-mail: athulpg007@gmail.com

## I. INTRODUCTION

The study of planetary atmospheres is critical to our understanding of the origin and evolution of the Solar System [1]. The combined effect of various physical and chemical processes over billions of years have resulted in a variety of planetary atmospheres across the Solar System as shown in Figure 1. The terrestrial planets Venus, Earth, and Mars have comparatively short atmospheres extending up to about 100 km above the surface, whereas the outer planets Jupiter, Saturn, Uranus, Neptune, and Saturn's moon Titan have their bigger atmospheres extend to about 1000 km. In addition to their scientific value, the diversity of atmospheres also has engineering implications for planetary missions [2]. For example, all landed missions at Venus, Mars, and Titan have made use of the atmosphere to slow down landers. The use of aerocapture, which uses a pass through the atmosphere to achieve spacecraft orbit insertion has also been extensively studied at all Solar System destinations [3, 4]. This paper performs a comparative study of the planetary atmospheres and their engineering implications for future entry and aerocapture missions.

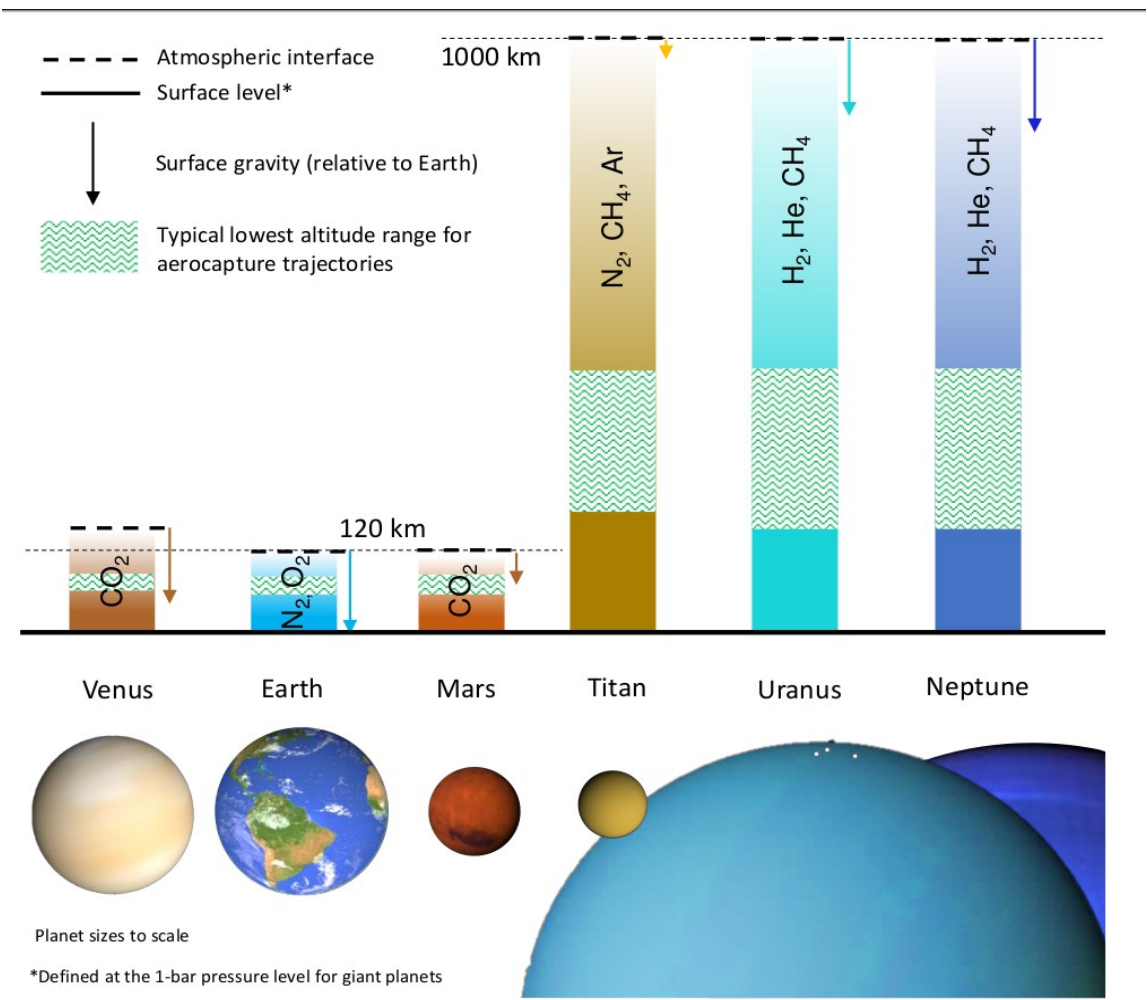

Figure 1. Schematic showing the variety of atmospheres in the Solar System.

## II. ATMOSPHERIC STRUCTURE AND COMPOSITION

Earth and Venus have similar mass and radius leading to very similar gravity and atmospheric extent, but the Venus atmosphere is much thicker than that of Earth, approximately by two orders of magnitude as seen in Figure 2. Mars however has only about 40% of Earth's gravity and the density at the surface is nearly two orders of magnitude less compared to that of Earth. The thicker Venus atmosphere implies aerodynamic heating rates for atmospheric entry are much higher at Venus compared to Earth, even though the entry speeds are similar due to identical gravity. Figure 3 shows the pressure profiles which are similar to the density profiles. The pressure at the Venusian surface is close to two orders of magnitude (92 bar) higher compared to that on Earth making it an extreme environment for lander missions, while the pressure on the Martian surface is two orders of magnitude less than that on Earth.

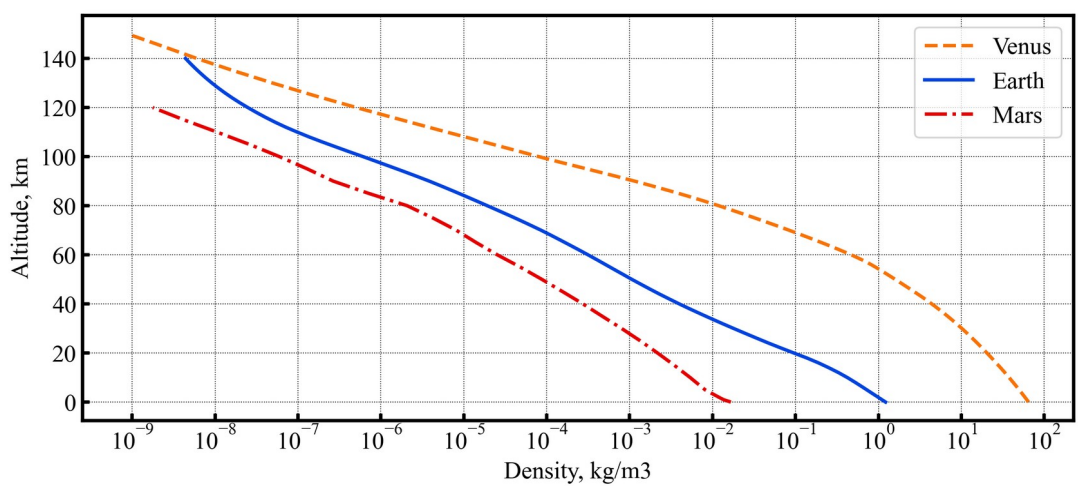

Figure 2. Density profiles for Venus, Earth, and Mars.

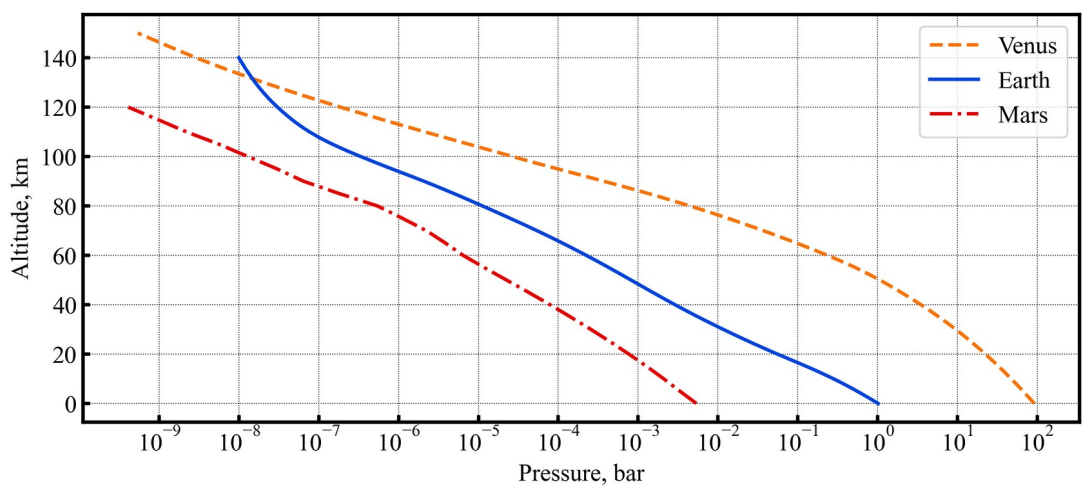

Figure 3. Pressure profiles for Venus, Earth, and Mars.

Figure 4 shows the temperature profiles for Venus, Earth, and Mars. The high temperature of the lower Venusian atmosphere due to the runaway greenhouse effect is prominent, making its surface temperature nearly 400 K hotter than that of Earth. However, higher up in the atmosphere at about 55-60 km, both the temperature and pressure are similar to that on Earth's surface. This makes the Venusian upper atmosphere a quite benign environment (discounting the corrosive sulfuric acid clouds). In fact, the significantly more benign conditions compared to the surface along with the much thicker atmosphere makes it an ideal choice for floating aerial platforms such as balloons [5].

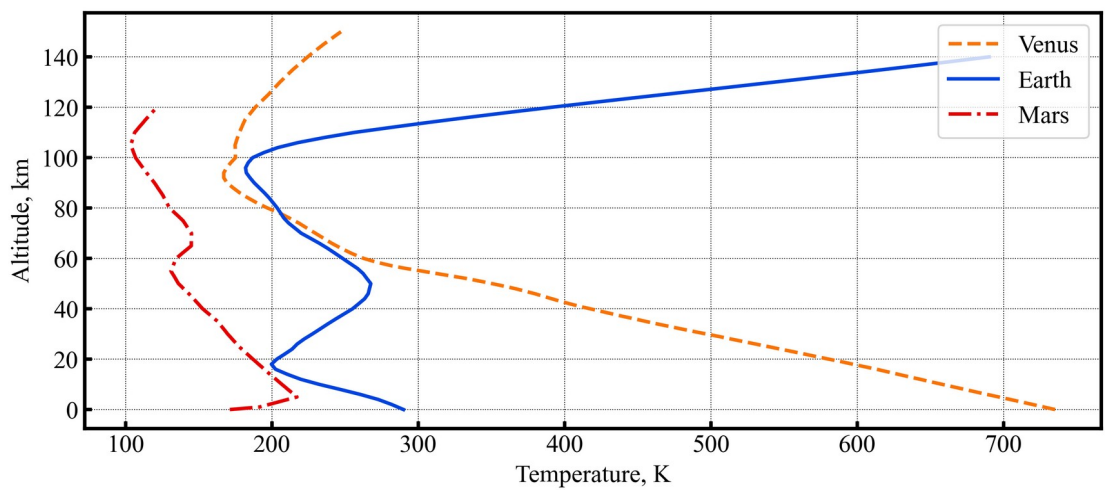

Figure 4. Temperature profiles for Venus, Earth, and Mars.

Figure 5 shows the density profiles for the outer planets and Titan. Compared to the inner planets, the outer planets have atmospheres that extend to nearly 1000 km above their surface (defined at 1 bar pressure level for giant planets). Their high gravity well causes much higher entry speeds and heating than encountered at the inner planets.

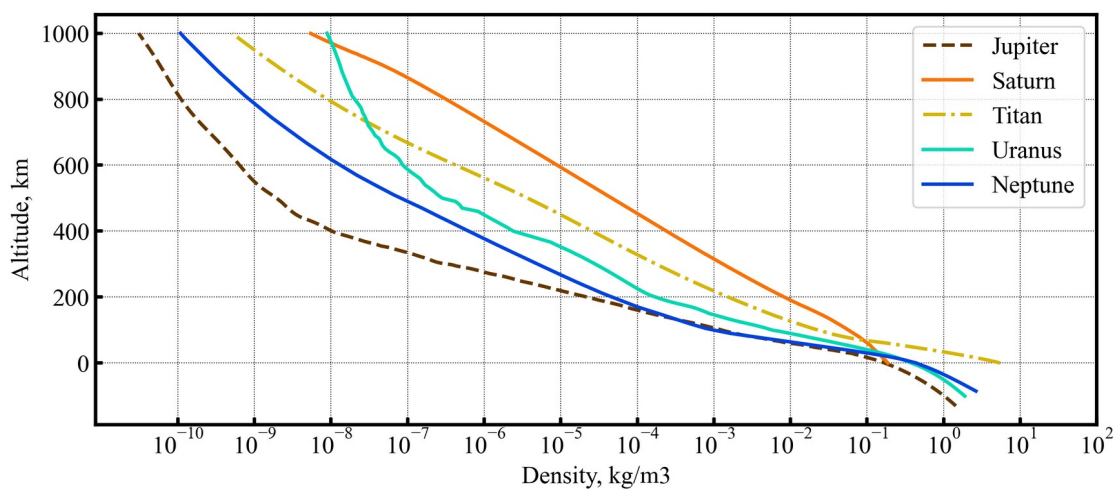

Figure 5. Density profiles for the outer planets and Titan.

Titan is one exception in the outer solar system with its low-gravity and the greatly extended thick atmosphere, which results in the low entry speed and low aerodynamic heating. Figures 6 and 7 show the pressure and temperature profiles. Titan's surface pressure is approximately 1.5 bar, with a temperature of 94 K leads to a density which is close to four times that on Earth. The low gravity and the dense atmosphere make Titan an ideal location for flying vehicles, which is exploited by the Dragonfly mission [6]. Titan's extended atmosphere also enables a long descent time of well over one hour, simplifying the atmospheric entry operations. Titan's atmospheric temperature profile, even though much colder, shows a striking comparison to that of Earth as seen in Figure 8. The temperature drops with altitude, and much like the ozone layer causes warming of Earth's upper atmosphere, the haze and smog layers cause warming on Titan. The ice giant atmospheres have the coldest temperatures in the Solar System, reaching close to 50 K.

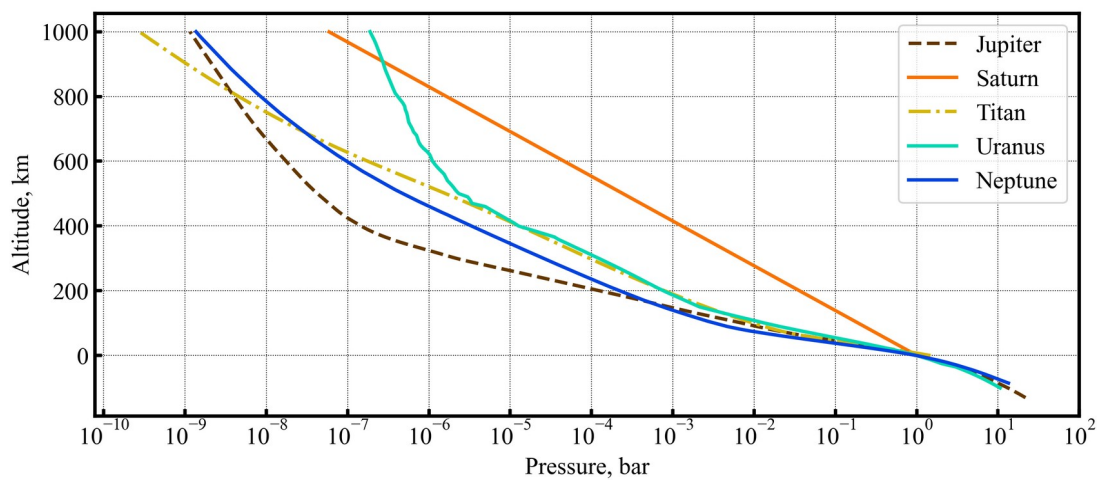

Figure 6. Pressure profiles for the outer planets and Titan.

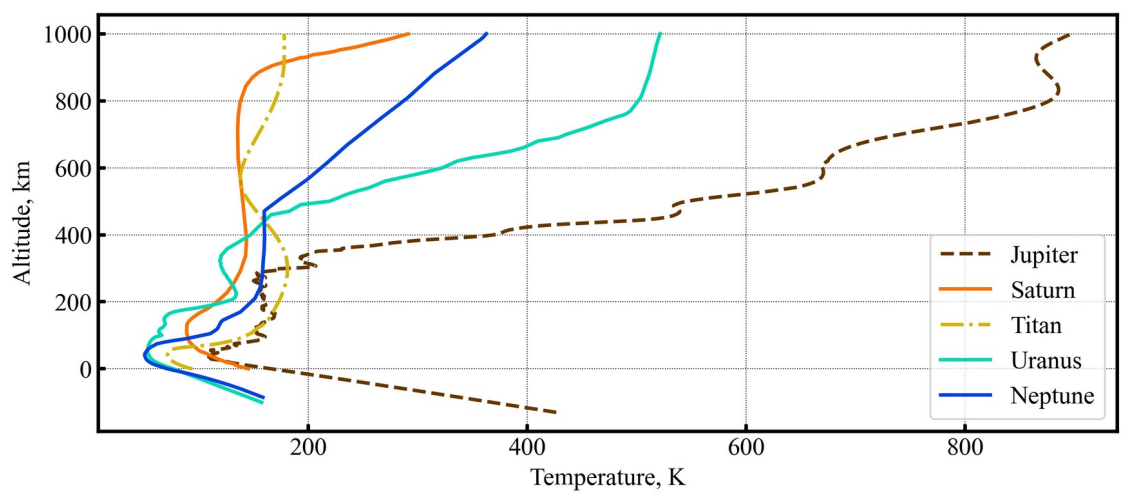

Figure 7. Temperature profiles for the outer planets and Titan.

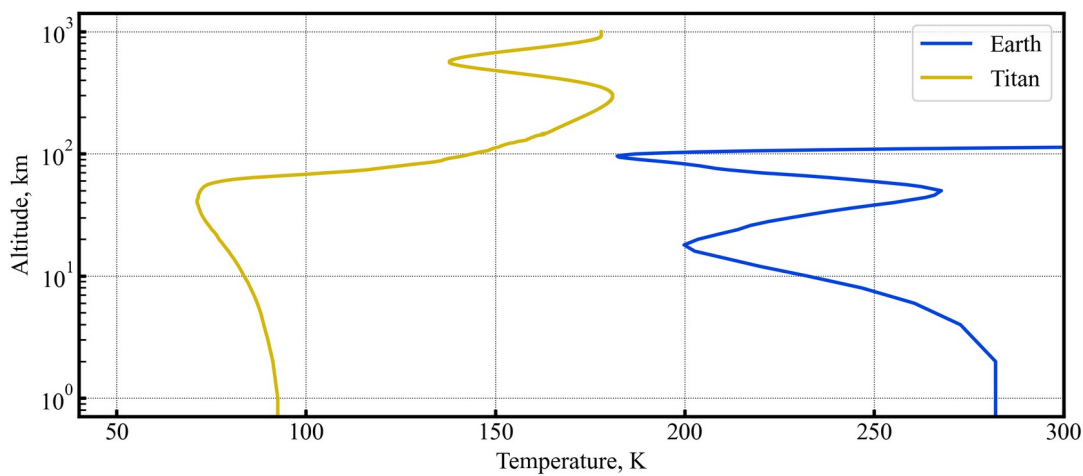

Figure 8. Comparison on Earth and Titan temperature profiles.

Figure 9 shows a comparison of the chemical composition of the planetary atmospheres at the surface level. Venus and Mars are primarily $CO_2$ dominated, with a small percentage of $N_2$. The Earth's atmosphere is primarily $N_2$, with the rest being mostly $O_2$, a product of biological activity. While most of the Earth's carbon is locked away in rocks, Mars and Venus must have had past geological process release most of the carbon into the atmospheres we see today. $CO_2$ being a greenhouse gas shows the effect of an extreme case of runaway global warming on Venus. Jupiter and Saturn atmospheres are primarily $H_2$, He consistent with solar proportions. Titan is the only other body which closely resembles the Earth's atmosphere in composition, supporting the hypothesis that Titan's atmosphere today may resemble that of the early Earth, and likely Venus and Mars [7]. Titan has a significant percentage of methane in its atmosphere, which also being a greenhouse gas provides some warming. In addition, Titan's temperature of 94 K being near the triple point of methane enables surface lakes, an active hydrological cycle, and weather patterns [8].

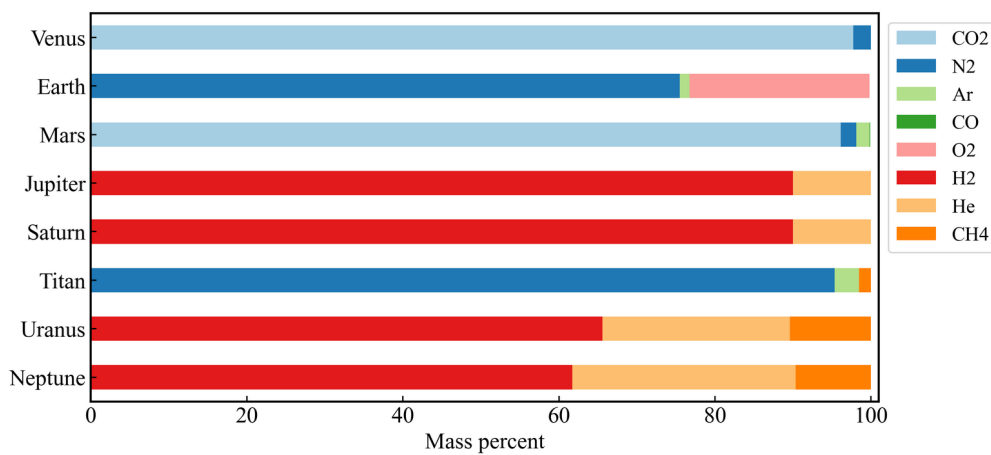

Figure 9. Comparison of the atmospheric composition (at the surface).

## III. PROBE FREE-FALL EXPERIMENT

All planetary entry missions use the deceleration provided by the atmosphere to slow down spacecraft. In this section, we perform a numerical experiment of dropping a descent probe from the top of the atmosphere at each planet and analyze its descent profile. The hypothetical descent probe is modeled based on the Galileo probe, with a mass of 125 kg, and a diameter of 0.66 m. The probe is dropped from the top of the atmosphere and its trajectory is propagated until surface impact using AMAT [9]. The key observable parameters are the descent time to the surface, altitude at which deceleration begins, maximum speed achieved, and vehicle speed at impact. This simple free-fall experiment provides valuable insights into the comparative deceleration effects provided by the various planetary atmospheres.

Figure 10 shows the comparison of the descent time of the probe (until surface impact) on Venus, Earth, and Mars. The effect of the thick Venusian atmosphere is prominent, taking the probe over 45 minutes to reach the surface. The hypothetical probe starts experiencing the deceleration effect at about 60 km above the surface, and it slows down considerably even without a parachute. As the probe descents, the increasing density slows it down more the further it descends into the atmosphere. In fact the Venusian lower atmosphere is so thick, that it is comparable to the probe falling in an ocean than an atmosphere. On the Earth, the descent rate is initially comparable to that of Venus, due to the similar gravity but the Earth's atmosphere being not as thick as Venus, the probe starts decelerating only below 20 km. The descent time is approximately 5 minutes. On Mars, the initial descent rate is slower due to the low gravity, but the total descent time is again 5 minutes. The thin Martian atmosphere is unable to effect any significant deceleration on the probe, as seen in the slope of the descent profile still being quite high at impact compared to Earth and Venus.

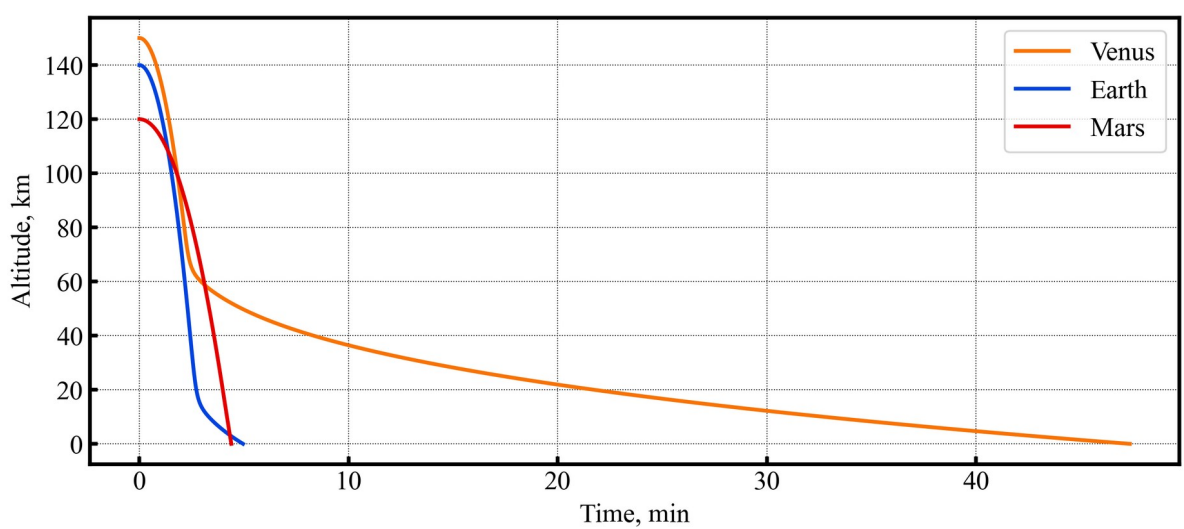

Figure 10. Comparison of free-fall descent profile to the surface on Venus, Earth, and Mars.

Figure 11 shows the free-fall speed profiles for Venus, Earth, and Mars. On Venus, the maximum speed achieved is about 1025 m/s before the dense atmosphere begins to decelerate the probe. On the Earth, the probe continues to accelerate further reaching about 1300 m/s before decelerating. On Mars, the probe accelerates but slowly due to its lower gravity, reaching a speed of 820 m/s and deceleration only begins to set in shortly before impact. The impact speed is highest at Mars, at close to 770 m/s as the thin atmosphere has not had enough time to act on the probe to slow it down. On the Earth, the impact speed is 78 m/s. On Venus, the impact speed is 10 m/s, as the thick atmosphere has slowed down the probe considerably, slow enough for a landing even without the need of a parachute. Figure 12 shows the comparison of the descent profiles at the outer planets and Titan. Titan's low gravity and thick extended atmosphere results in a unusually long descent time of well over an hour, the longest of any Solar System destination.

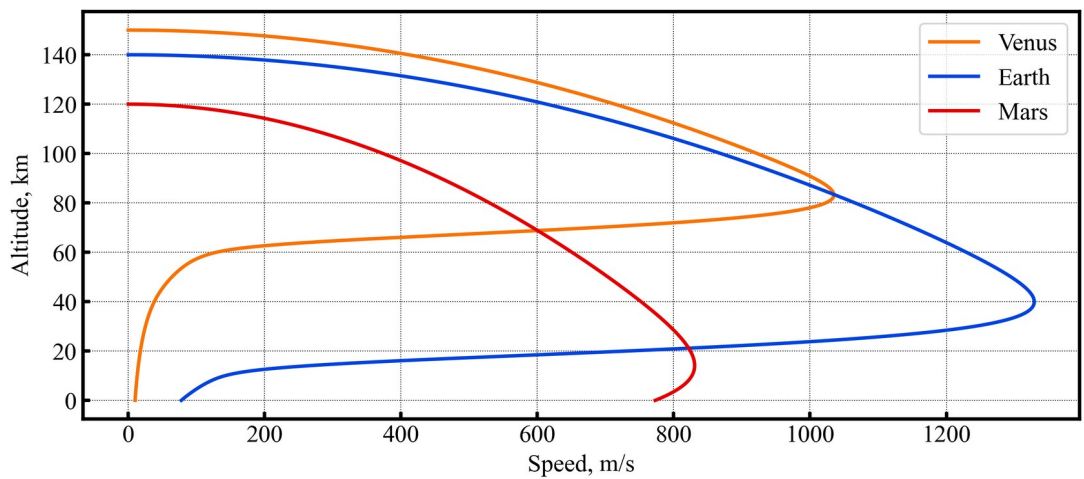

Figure 11. Comparison of free-fall descent speed profile on Venus, Earth, and Mars.

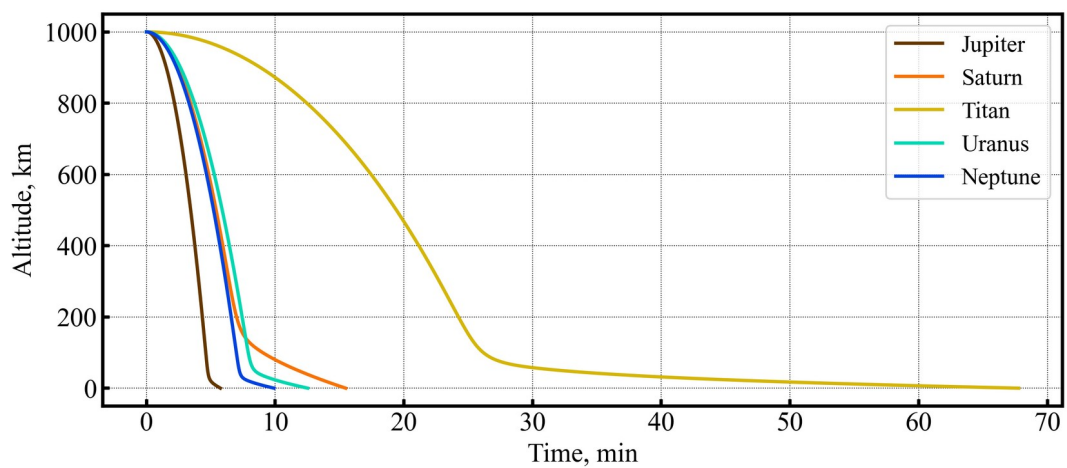

Figure 12. Comparison of free-fall descent profile to the surface on outer planets and Titan.

Figure 13 shows the free-fall speed profiles on the outer planets and Titan. On Jupiter with its enormous gravity well, the hypothetical probe reaches speeds of over 6000 m/s before the atmosphere begins decelerating it. On the other hand, Titan has the smallest gravity well, and the probe reaches a maximum speed of 1000 m/s. On Titan, like Venus the dense lower atmosphere results in a low impact speed of 13 m/s, even without a parachute.

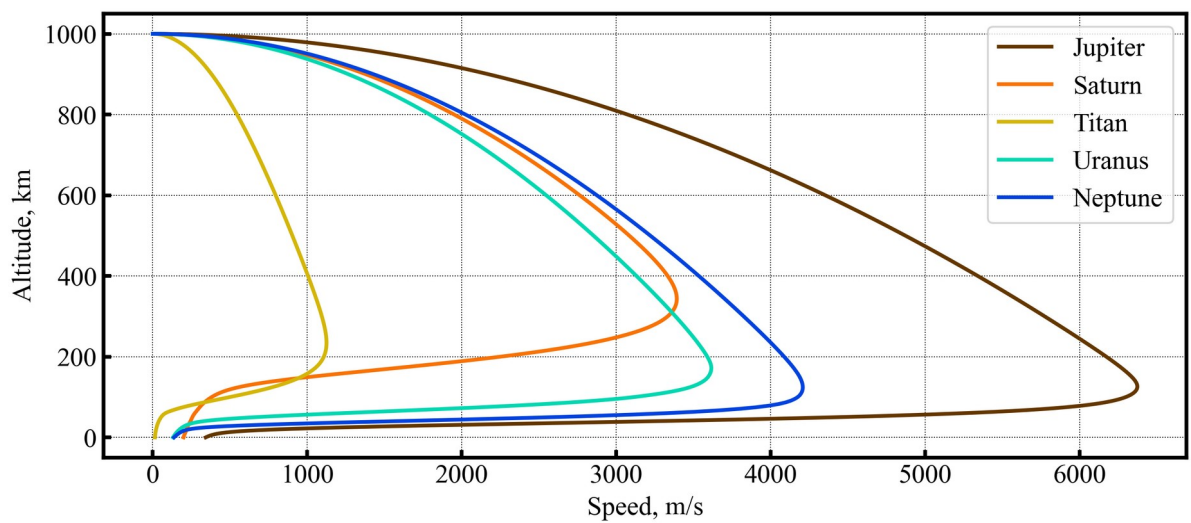

Figure 13. Comparison of free-fall speed profile on the outer planets and Titan.

### IV. AERODYNAMIC DECELERATION AND HEATING

Figure 14 shows the comparison of entry deceleration profiles at Venus, Earth, and Mars. The entry speed, flight-path angle is [11 km/s, -10 deg] at Earth and Venus, and [6 km/s, -15 deg.] at Mars. The effect of the thick Venusian is prominent, as the Venus atmosphere causes much higher deceleration than at Earth (75g vs 36g) to occur at altitudes as high as 80 km. For Mars entry, the deceleration is only 12g and occurs at an altitude of about 10 km.

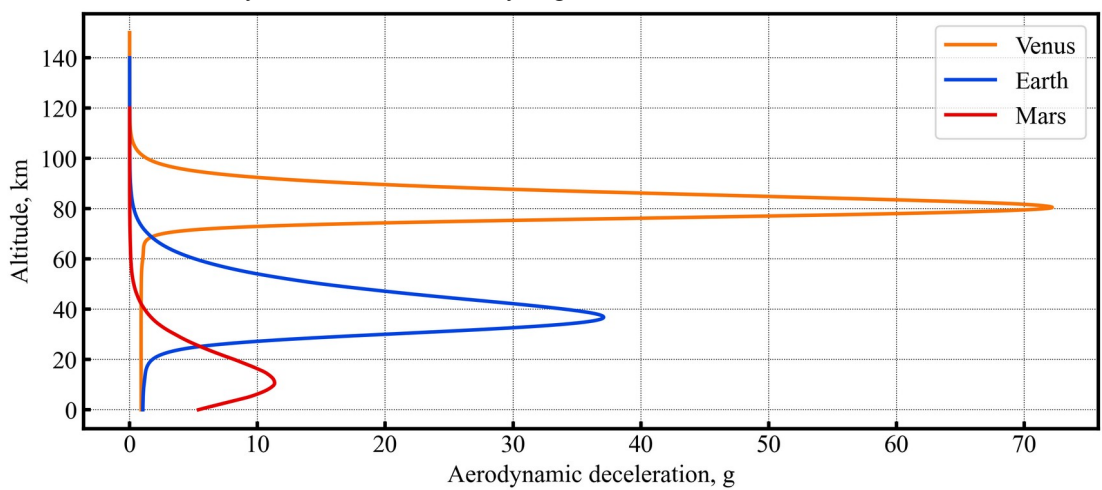

Figure 14. Comparison of entry deceleration profiles on Venus, Earth, and Mars.

Figure 15 shows the comparison of entry aerodynamic heating profiles at Venus, Earth, and Mars. Once again the thick Venusian atmosphere results in the highest aerodynamic heating, close to 2500 W/cm². The presence of $CO_2$ in the Venusian atmosphere is a contributor to the significantly higher radiative heating at Venus than Earth. The thin Martian atmosphere results in considerably less heating under 200 W/cm², an order of magnitude less than Venus.

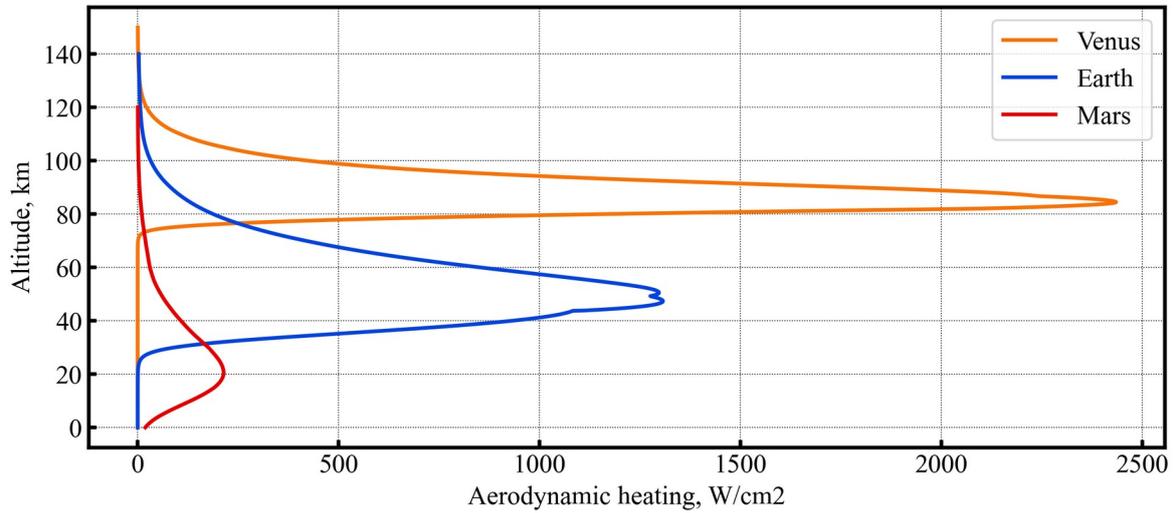

Figure 15. Comparison of entry aerodynamic heating profiles on Venus, Earth, and Mars.

Figure 16 shows the comparison of entry deceleration profiles at the outer planets and Titan. The entry speed, flight-path angle are: Jupiter [50 km/s, -10 deg], Saturn [30 km/s, -10 deg], Titan [7 km/s -50 deg], Uranus and Neptune: [23 km/s, -15 deg.]. Jupiter entry results in the highest deceleration, while Titan is the most benign body,

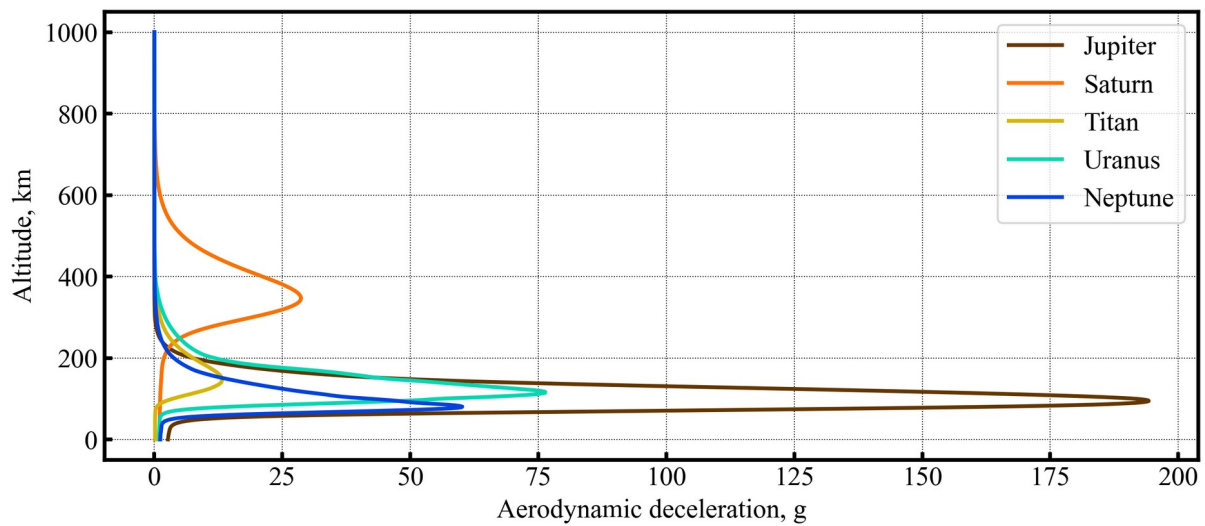

Figure 16. Comparison of entry aerodynamic deceleration at the outer planets.

Figure 17 shows the comparison of heating profiles at the outer planets and Titan. Jupiter presents the most extreme aero-thermal environment with heating rates in excess of 100,000 W/cm². Saturn, Uranus, and Neptune entries encounter heating in the range of 1000-5000 W/cm², which is much higher than that at Earth or Mars, but comparable to Venus. Titan presents the most benign aero-thermal environment for atmospheric entry anywhere in the Solar System. Figure 18 shows a comparison of the entry aero-thermal environments across the Solar System, ranging from Titan and Mars which are benign targets, to Jupiter which is the most extreme.

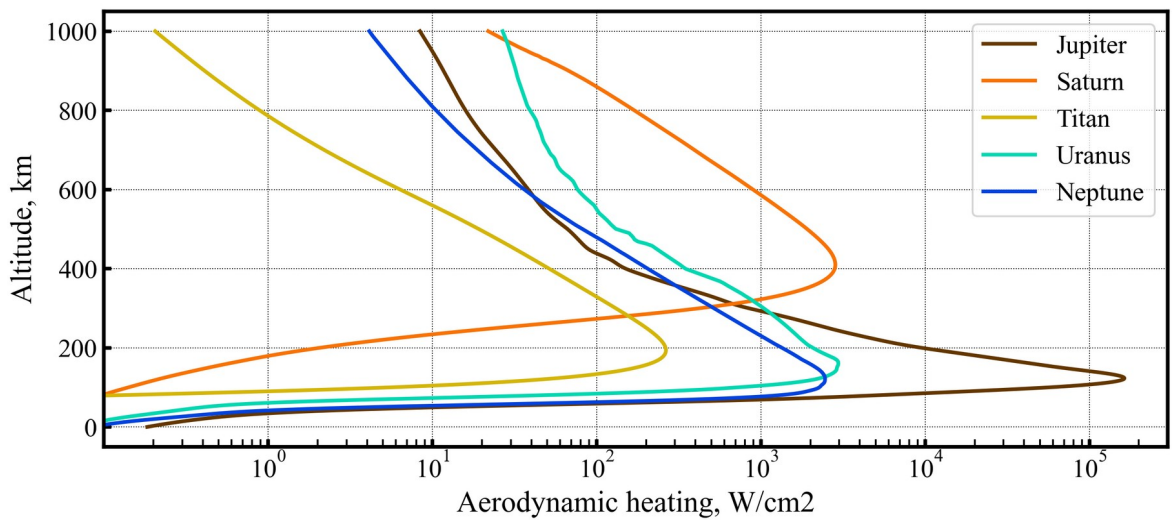

Figure 17. Comparison of entry aerodynamic heating profiles at the outer planets.

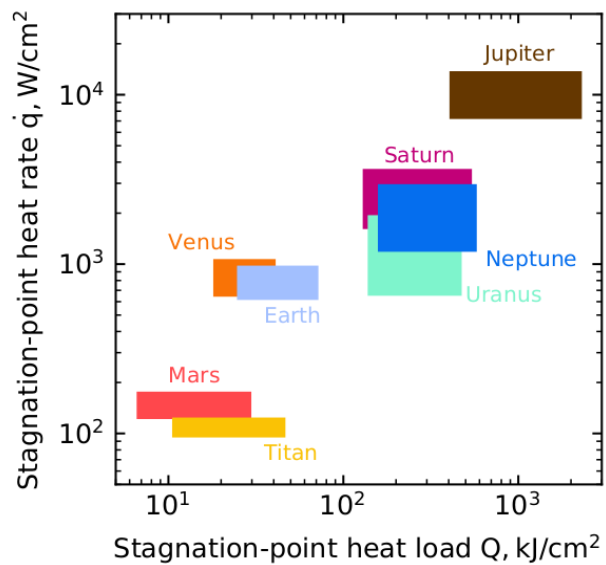

Figure 18. Comparison of entry aero-thermal environments in the Solar System.

## V. IMPLICATIONS FOR FUTURE MISSIONS

Among the inner planets, Venus has the thickest atmosphere and the highest surface temperature and pressure. The high deceleration and aerodynamic heating rates at Venus present a more demanding entry environment than the Earth or Mars. The heating rates are within the capability of ablative thermal protection systems such as HEEET and will likely be used on future entry probe missions. Aerocapture at Venus using rigid aeroshells is technically feasible, but not an attractive option due to the high deceleration and heating rates [10]. However, aerocapture is an attractive option for inserting small satellites into orbit using deployable low-ballistic coefficient systems. By decelerating much higher up in the atmosphere where the density is lower, the deployable entry systems can keep the heating rates much lower than rigid aeroshells [11]. This has potential applications to sending small satellites to Venus orbit as independent secondary payloads [12], supporting payloads for a larger mission [13], and sample return missions [14].

Mars has a fairly thin atmosphere which can provide some deceleration for entry, but is often not enough to slow them down enough before they reach the surface. With the lowest deceleration and heating rates in the inner Solar System, Mars presents a relatively benign target for entry and aerocapture. This makes it particularly attractive for payloads using deployable entry systems [15], and a low-cost aerocapture technology demonstrator mission [16].

Jupiter and Saturn have enormous gravity wells which result in very high entry speeds, and results in the most extreme deceleration and aero-thermal loads anywhere in the Solar System. This presents severe challenges for atmospheric probe entry and makes orbiter aerocapture practically infeasible at these destinations [17].

Titan is a unique destination, with its low gravity and extended thick atmosphere making it the most benign destination for both planetary entry and aerocapture. Its low gravity implies the entry speeds are the lowest anywhere in the outer Solar System, and the thick extended atmosphere keeps the deceleration and heating rates low. The long descent time to the surface which takes over an hour simplifies entry operations compared to Mars entry which takes only a few minutes. For aerocapture, the combination of low gravity and extended atmosphere results in the largest scale height, and consequently provides the largest corridor width anywhere in the Solar System [18]. This has applications for a landers delivered to Titan surface using deployable systems, as well as a future Titan orbiter [19].

Uranus and Neptune also have large gravity wells, resulting in high entry speeds and high aerodynamic heating, but they are much less demanding than Jupiter or Saturn. Typical heating rates are in the range of 1000-5000 W/cm$^2$, within the capability of HEEET thermal protection system [20]. While aerocapture is not considered for the currently proposed Uranus Orbiter and Probe Flagship mission [21, 22], aerocapture can enable significantly shorter trip times, as well as insert more useful mass into orbit around both Uranus and Neptune for future missions [23, 24, 25].

## VI. CONCLUSIONS

The combined effect of various physical and chemical processes over billions of years have resulted in a variety of planetary atmospheres across the Solar System. In addition to their scientific value, the diversity of atmospheres also has engineering implications for planetary missions. The paper performed a comparative study of the planetary atmospheres and explored their engineering implications for future entry and aerocapture missions. The atmospheric structure and composition of the various planetary bodies is compared. The thick Venusian atmosphere results in high deceleration and heating rates and presents a demanding environment for both atmospheric entry and aerocapture. However, aerocapture is an attractive option for inserting small satellites into orbit using deployable low-ballistic coefficient system. The thin Martian atmosphere allows low aerodynamic heating, but itself is not enough to decelerate a lander to sufficiently low speeds for a soft landing. With the lowest deceleration and heating rates in the inner Solar System, Mars presents a relatively benign target for entry and aerocapture missions. This makes it particularly attractive for payloads using deployable entry systems, and a low-cost aerocapture technology demonstrator mission. With their enormous gravity wells, Jupiter and Saturn entry result in very high entry speeds, deceleration, and heating making them the most demanding destinations for atmospheric entry and impractical for aerocapture. Saturn's moon Titan is a unique destination in the Solar System, with its low gravity and greatly extended thick atmosphere. Its low gravity implies the entry speeds are the lowest anywhere in the outer Solar System, and the thick extended atmosphere keeps the deceleration and heating rates low. The long descent time to the surface which takes over an hour simplifies entry operations compared to Mars entry which takes only a few minutes. The combination of low gravity and extended atmosphere results in the largest scale height, and consequently provides the largest corridor width and most benign environment anywhere in the Solar System for both atmospheric entry and aerocapture. Uranus and Neptune also have large gravity wells, resulting in high entry speeds and high aerodynamic heating, but they are much less demanding than Jupiter or Saturn. Existing entry vehicles and thermal protection system materials are sufficient for both atmospheric entry and aerocapture at Uranus and Neptune. While aerocapture is not considered for the currently proposed Uranus Orbiter and Probe Flagship mission, aerocapture has been shown to be able to enable significantly shorter trip times, as well as insert more useful mass into orbit around both Uranus and Neptune.

## DATA AVAILABILITY

The atmospheric models were generated using NASA Global Reference Atmospheric Model (GRAM) Suite v1.5. The probe trajectories were propagated using the open-source Aerocapture Mission Analysis Tool (AMAT) v2.2.22. The data presented in the paper will be made available by the author upon reasonable request.

# REFERENCES

[1] National Academies of Sciences, Engineering, and Medicine. 2022. *Origins, Worlds, and Life: A Decadal Strategy for Planetary Science and Astrobiology 2023-2032*. Washington, DC: The National Academies Press.
https://doi.org/10.17226/26522

[2] Girija AP, "A Systems Framework and Analysis Tool for Rapid Conceptual Design of Aerocapture Missions," Ph.D. Dissertation, Purdue University Graduate School, 2021.
https://doi.org/10.25394/PGS.14903349.v1

[3] Spilker T et al., "Qualitative assessment of aerocapture and applications to future missions," *Journal of Spacecraft and Rockets,* Vol. 56, No. 2, 2019, pp. 536-545.
https://doi.org/10.2514/1.A34056

[4] Girija AP et al. "Quantitative assessment of aerocapture and applications to future solar system exploration." *Journal of Spacecraft and Rockets,* Vol. 59, No. 4, 2022, pp. 1074-1095.
https://doi.org/10.2514/1.A35214

[5] Limaye SS, and Garvin JB, "Exploring Venus: next generation missions beyond those currently planned," *Frontiers in Astronomy and Space Sciences,* Vol. 10, 2023, pp. 1188096.
https://doi.org/10.3389/fspas.2023.1188096

[6] Barnes, JW et al. "Science goals and objectives for the Dragonfly Titan rotorcraft relocatable lander," *The Planetary Science Journal,* Vol. 2, No. 4, 2021, pp. 130.
https://doi.org/10.3847/PSJ/abfdcf

[7] He C, Smith MA, "Identification of nitrogenous organic species in Titan aerosols analogs: Implication for prebiotic chemistry on Titan and early Earth," *Icarus*, Vol. 238, 2014, pp. 86-92.
https://doi.org/10.1016/j.icarus.2014.05.012

[8] Hayes AG et al., "A post-Cassini view of Titan's methane-based hydrologic cycle," *Nature Geoscience,* Vol. 11, No. 5, 2018, pp. 306-313.
https://doi.org/10.1038/s41561-018-0103-y

[9] Girija AP et al. "AMAT: A Python package for rapid conceptual design of aerocapture and atmospheric Entry, Descent, and Landing (EDL) missions in a Jupyter environment," *Journal of Open Source Software,* Vol. 6, No. 67, 2021, pp. 3710.
https://doi.org/10.21105/joss.03710

[10] Girija AP, Lu Y, and Saikia SJ, "Feasibility and mass-benefit analysis of aerocapture for missions to Venus," *Journal of Spacecraft and Rockets,* Vol. 57, No. 1, 2020, pp. 58-73.
https://doi.org/10.2514/1.A34529

[11] Dutta S et al., "Mission Sizing and Trade Studies for Low Ballistic Coefficient Entry Systems to Venus," *2012 IEEE Aerospace Conference,* IEEE, 2012, pp. 1-14.
https://doi.org/10.1109/AERO.2012.6187002

[12] Girija AP, Saikia SJ, and Longuski JM, "Aerocapture: Enabling Small Spacecraft Direct Access to Low-Circular Orbits for Planetary Constellations," *Aerospace,* Vol. 10, No. 3, 2023, pp. 271.
https://doi.org/10.3390/aerospace10030271

[13] Limaye SS et al., "Venus observing system," *Bulletin of the American Astronomical Society,* Vol. 53, No. 4, 2021, pp. 370.
https://doi.org/10.3847/25c2cfeb.7e1b0bf9

[14] Shibata E et al., "A Venus Atmosphere Sample Return Mission Concept: Feasibility and Technology Requirements," *Planetary Science Vision 2050 Workshop,* 2017, pp. 8164.

[15] Austin A et al., "Enabling and Enhancing Science Exploration Across the Solar System: Aerocapture Technology for SmallSat to Flagship Missions," *Bulletin of the American Astronomical Society,* Vol. 53, No. 4, 2021, pp. 057.
https://doi.org/10.3847/25c2cfeb.4b23741d

[16] Girija AP, "A Low Cost Mars Aerocapture Technology Demonstrator," arXiv, Vol. 2307, No. 11378, 2023, pp. 1-14
https://doi.org/10.48550/arXiv.2307.11378

[17] Girija AP, "Aerocapture: A Historical Review and Bibliometric Data Analysis from 1980 to 2023," arXiv, Vol. 2307, No. 01437, 2023, pp 1-19.
https://doi.org/10.48550/arXiv.2307.01437

[18] Girija AP et al., "A Unified Framework for Aerocapture Systems Analysis," *AAS/AIAA Astrodynamics Specialist Conference,* 2019, pp 1-21.
https://doi.org/10.31224/osf.io/xtacw

[19] Girija AP, "ADEPT Drag Modulation Aerocapture: Applications for Future Titan Exploration," arXiv, Vol. 2306, No. 10412, 2023, pp 1-27.
https://doi.org/10.48550/arXiv.2306.10412

[20] Girija AP, et al., "Feasibility and performance analysis of neptune aerocapture using heritage blunt-body aeroshells," *Journal of Spacecraft and Rockets,* Vol. 57, No. 6, 2020, pp. 1186-1203.
https://doi.org/10.2514/1.A34719

[21] Jarmak S et al., "QUEST: A New Frontiers Uranus orbiter mission concept study," *Acta Astronautica,* Vol. 170, 2020, pp. 6-26.
https://doi.org/10.1016/j.actaastro.2020.01.030

[22] Cohen I et al., "New Frontiers-class Uranus Orbiter: Exploring the feasibility of achieving multidisciplinary science with a mid-scale mission," *Bulletin of the American Astronomical Society,* Vol. 53, No. 4, 2021, pp. 323.
https://doi.org/10.3847/25c2cfeb.262fe20d

[23] Dutta S et al., "Aerocapture as an Enhancing Option for Ice Giants Missions," *Bulletin of the American Astronomical Society,* Vol. 53, No. 4, 2021, pp. 046.
https://doi.org/10.3847/25c2cfeb.e8e49d0e

[24] Girija AP, "A Flagship-class Uranus Orbiter and Probe mission concept using aerocapture," *Acta Astronautica* Vol. 202, 2023, pp. 104-118.
https://doi.org/10.1016/j.actaastro.2022.10.005

[25] Iorio L et al., "One EURO for Uranus: the Elliptical Uranian Relativity Orbiter mission." *Monthly Notices of the Royal Astronomical Society*, Vol. 523, No. 3, 2023, pp. 3595-3614
https://doi.org/10.1093/mnras/stad1446